\documentstyle[prl,aps,epsfig]{revtex}

\begin{document}

\draft

\title{Dynamical mechanisms of DC current generation in 
driven Hamiltonian systems}

\author{S. Denisov and S. Flach}
\address{Max-Planck-Institut f\"ur Physik komplexer Systeme,
N\"othnitzer Str. 38, D-01187 Dresden, Germany}

\date{\today}

\wideabs{

\maketitle

\begin{abstract}
Recent symmetry considerations (Phys. Rev. Lett. {\bf 84} 2358 (2000))
have shown that dc currents may be generated in the stochastic layer
of a system describing the motion of a particle in a one-dimensional
potential in the presence of an ac time-periodic drive. In this paper
we explain the dynamical origin of this current.
We show that the dc current is induced  
by the presence and desymmetrization of ballistic channels
inside the stochastic layer. The existence of these 
channels is due to resonance islands with 
non-zero winding numbers. The characterization of the
flights dynamics inside ballistic channels is described
by distribution functions. We obtain these distribution functions
numerically and find very good agreement with simulation data.
\end{abstract}


}

Transport in driven systems has received widespread interest 
for several years because of its potential applicability to
nonequilibrium processes \cite{fjaajp97}. One canonical model reduces to the
motion of a particle in a one-dimensional space-periodic potential
in the presence of friction and a time-dependent stochastic force 
$\chi(t)$ . 
If $\chi(t)$ contains correlations, a nonzero current
may be realized even in the case of zero average $< \chi(t) > =0$ 
\cite{fjaajp97}.
Despite an enormous accumulation of results in this area
\cite{pr01} we are still lacking
a full understanding of the {\it microscopic} mechanisms of current occurrence.
If such an understanding is realizable, it should make use of the 
true dynamical evolution of the system rather than of properties of
equations for probabilities. A first step in such a direction 
requires to separate the essential time correlations from the pure
Gaussian white noise in $\chi(t)$. The simplest way is to assume
that $\chi(t)=E(t)+\xi(t)$ where $E(t)=E(t+T)$ is a time-periodic function
with zero mean $<E(t)>=0$ and $\xi(t)$ is a Gaussian white noise term.

The next step is to skip the $\xi(t)$ term which leaves us with
a regular dynamical problem. In \cite{sfoyyz00} such a case was
considered and the relevant space-time symmetries of the dynamical problem
have been obtained. It was shown that a breaking of those symmetries
leads to a nonzero dc current. The mechanism of current occurrence 
for the dissipative case was identified with a desymmetrization
of attractor basins. In contrast the nondissipative (Hamiltonian)
case is much less understood \cite{igph00}. 
A time-dependent Hamiltonian system is usually
nonintegrable \cite{Licht}. 
A strong dc current component was found in the corresponding
stochastic layer of such a system \cite{sfoyyz00}. 
While its presence or absence
was clearly connected to the above mentioned absence or presence of 
symmetries, the dynamical nature of directed transport in the
stochastic layer lacks full understanding. 
The importance of this understanding
can be seen from e.g. results in \cite{oysfaaoyz01} where kinetic equations
for probability functions have been studied. In particular it was
found that the approaching of the Hamiltonian (dissipationless) limit
leads to an increase of the dc current value by 2-3 orders of magnitude.
Thus the description of the dynamical 
mechanisms of directed current generation in
the stochastic layer of a driven Hamiltonian system will provide with
very useful information for dissipative systems as well.

Let us consider the canonical example of a
particle moving in a spatially periodic nonlinear potential $U(x)=-\cos x$
under
the influence of
time-periodic zero-mean force $E(t)$. The Hamiltonian and the 
equation of motion are
given by:
\begin{equation}\label{1}
H=\frac{p^2}{2} -\cos x - xE(t)\;\;,\;\;
\ddot{x} = -\sin x + E(t)\;\;.
\end{equation}
Here $p$ and $x$ are canonically conjugated momentum and coordinate
and $\ddot{x}\equiv {\rm d}^2 x / {\rm d} t^2$.

We restrict our consideration to the choice
\begin{equation} 
E(t)=E_1 \cos (t) + E_2 \cos (2t + \phi)\;\;.
\label{e(t)}
\end{equation}
According to \cite{sfoyyz00} for $E_2 \neq 0$ and $\phi \neq 0,\pi$
all possible symmetries which yield zero dc current are broken. 
Note that the phase space dimension $d$ of (\ref{1}) is $d=3$.

In the case of a nonzero field $E(t)$ 
the phase space of  
(\ref{1})
is characterized by the presence of a stochastic layer 
which originates from the
destroyed separatrix of the undriven system \cite{Licht}. 
For $\phi=0,\pi$ this layer is invariant under the transformation 
($p \rightarrow -p$, $t \rightarrow -t$, $x \rightarrow x$). At the same
time the average velocity for any trajectory in this layer
vanishes, so we find zero dc current. 
The symmetry will be broken when 
tuning $\phi$ away from the values $(0,\pi)$. 
The stochastic layer will deform. Most importantly any trajectory in
the layer will then be characterized 
by a nonzero value of the average velocity.
Due to ergodicity inside the layer this value will be unique
for all trajectories from the layer. 
While the fact that it may become nonzero 
is understandable using the symmetry analysis,
its appearance and magnitude is due to dynamical mechanisms of motion inside
the stochastic layer.
In this paper, we show that the dc current is induced  
by the presence and desymmetrization of ballistic channels
inside the stochastic layer. The existence of these 
channels is due to resonances. The characterization of the
realization of flights inside ballistic channels is described
by distribution functions. We obtain these distribution functions
numerically and find very good agreement with simulation data.

System (\ref{1}) has a mixed phase space, which
contains of chaotic areas and regular resonance
islands \cite{Zas}. These islands are impermeable  for chaotic
trajectories and, at a first glance,
may be excluded from the consideration of 
the phase space flow inside the stochastic layer.
In reality the phase space topology inside the stochastic layer
is very complex
precisely at the boundary between chaotic and
regular
regions \cite{Zas}. Close to resonances the stochastic 
layer shows up with a hierarchical
set of
cantori, which form partial barriers for a trajectory from the
layer.
Due to the presence of these
barriers a chaotic trajectory can be trapped  
for a long time near a regular island resonance. This trapping or
sticking effect
leads to the appearance of strongly nonergodic
episodes during the overall chaotic motion. 
Regular islands are characterized by a corresponding
rational winding number $\omega=\Delta x / T$ which defines the distance 
$\Delta x$ travelled
during one period $T=2\pi$ of the drive.
If the winding number
$\omega$ is nonzero, the corresponding sticking episode
of the chaotic trajectory is a ballistic-like unidirectional
flight. For $\omega=0$ the sticking episode corresponds to trapped oscillations.

Thus the complicated evolution of a trajectory in the stochastic layer
can be subdivided into several parts \cite{Zas}. The first one is a fast
diffusion in the bulk of the layer, while the other ones are
stickings to the above mentioned regular islands and correspond
to propagation in
ballistic channels. The switching from the diffusion
process into a ballistic flight will be described by some
probability distribution. The same will be true for the actual
residence or sticking time inside a given channel.
We will show for the cases studied that the fast diffusion
alone is not capable of explaining the observed dc current.
The main point is, that the leading mechanism of current
generation in the {\it stochastic layer} is related to the 
desymmetrization of the strongly {\it non-stochastic} part
of the overall stochastic dynamics inside the layer. 
Note that our kinetic energy choice
$p^2/2$ in (\ref{1}) implies that the stochastic layer is bounded
in $p$, so we will always expect ballistic channels to appear.

Let us study the case of weak driving $E_1=0.252$ and $E_2=0.052$.
A Poincare map of the phase space flow for
$\phi=0$ is shown in Fig.1(a) \cite{com_poinc}.
The main stochastic layer (central location) shows up
with zero average velocity due to symmetry arguments (Fig.2). 
The large hole in the middle of this layer corresponds to
regular trapped motion in the well of $U(x)$.
Additional resonances are seen above and below
the central layer. These thin ballistic-like but yet stochastic channels
have no overlap with the central layer. 	
%%%%%%%%%%%%%%%%%%%%%%%%%%%%%%%%%%%%%%%%%%
%%%%%%%%%%%% FIG 1 %%%%%%%%%%%%%%%%%%%%%%%
%%%%%%%%%%%%%%%%%%%%%%%%%%%%%%%%%%%%%%%%%%
\begin{figure}[htb]
\vspace{10pt}
\centerline{
\epsfig{file=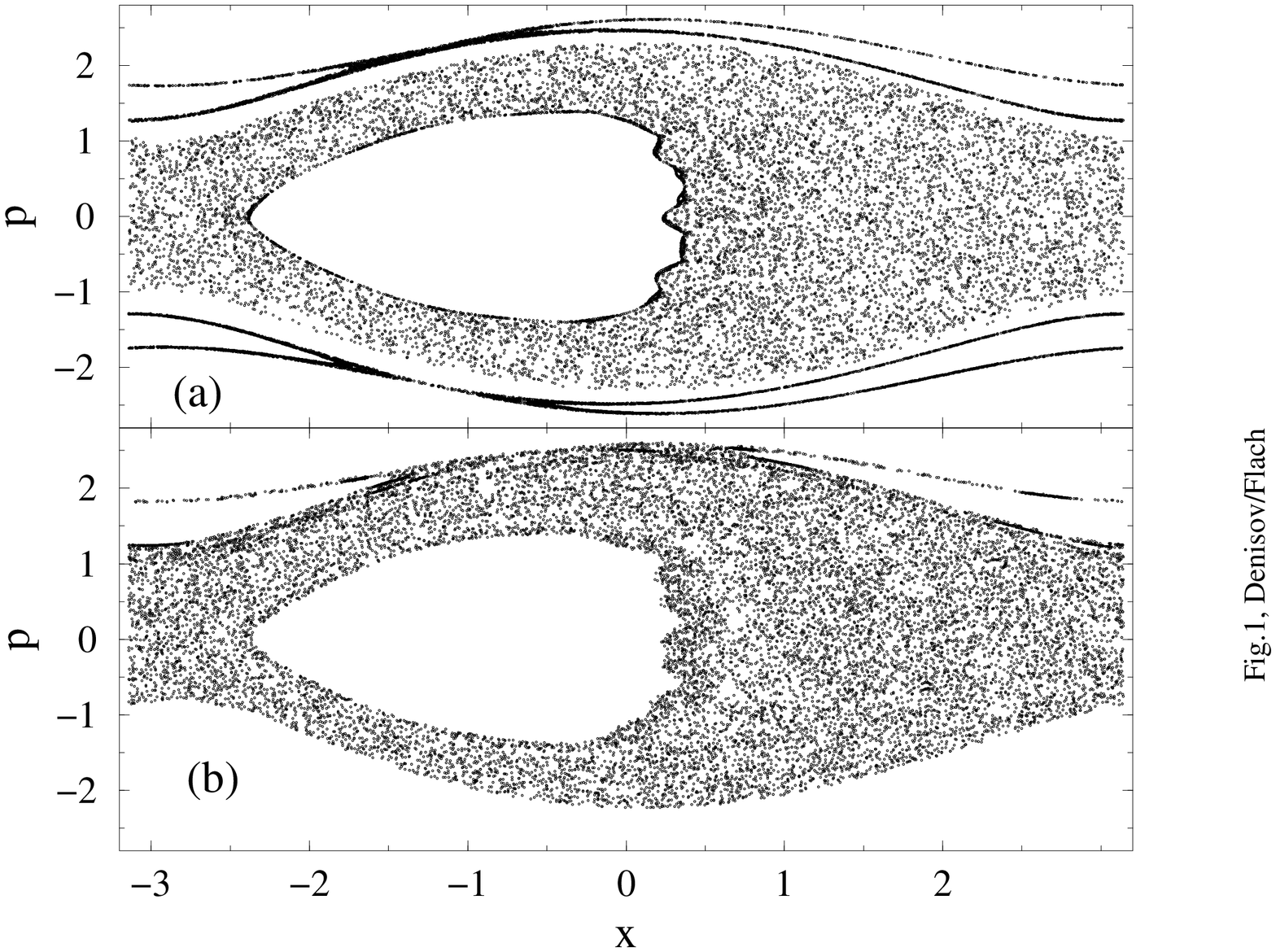,width=82mm,height=90mm,angle=270}
}
\vspace{2pt}
\caption{Poincare map for (a) $\phi=0$ and (b) $\phi=\pi/2$.}
\label{fig1}
\end{figure}
%%%%%%%%%%%%%%%%%%%%%%%%%%%%%%%%%%%%%%%%%%%

A weak  asymmetry $\phi=\pi/5$
leads to a slight deformation of the main stochastic layer and to a 
desymmetrization of the overlap of the chaotic layer with higher-order
resonances and to the appearance of a positive current in the system.  
Note that it still does not overlap with
the thin ballistic channels seen in Fig.1(a).
Most importantly we observe a nonzero average velocity 
$<\dot x> \approx 0.05$ (Fig.2).

Further increase
of the asymmetry, $\phi=\pi/2$,
results in an overlapping of the main stochastic
layer with the upper  ballistic resonance
Fig.1(b). Note that at the same time the lower ballistic resonance
is not overlapping.
The average velocity increases to $<\dot x > \approx 0.2$,
which is four times larger than the result for $\phi=\pi/5$.
Standard harmonic mixing theories (see \cite{pr01},\cite{igph98}) would predict a
dependence $< \dot x > \sim \sin \phi$ and thus only an increase by
a factor of 1.7. 

In the $x(t)$ curves in Fig.2 we observe many ballistic flights.
For one of them an inset shows the corresponding Poincare map result,
which verifies that these flights correspond to stickings of the
chaotic trajectory to the upper balistic resonances.
A zooming of the $x(t)$ curves shows {\it self-similarity},
i.e. the seemingly random dynamics between observable long flights
is actually again composed of shorter flights and  seemingly random dynamics
etc (cf. insets in Fig.2). 

In order to quantify our analysis of the symmetry broken dynamics
we compute the distribution of travelling times of
'uniform' flights  to the left $P_-(t_f)$ and
to the right $P_+(t_f)$ separately. Here 'uniform' means no change
of direction of motion \cite{com_num}. For each separate flight we note
both the time $t_f$ spent in this motion and the distance $x_f$ travelled. 
The dependence of $x_f$ on $t_f$ is shown in the inset of
Fig.3(a) for $\phi=\pi/2$. 
Similar to the other cases $\phi=0,\pi/5$ 
we observe a simple fork-like structure.
This is due to the fact that any considerable distance 
covering in the stochastic
layer is realized through flights while sticking to the boundary
of the stochastic layer. The slopes in the inset of
Fig.3(a) are given by the corresponding
winding numbers of the layer boundaries. Note that the two fork parts
merge at values of $t_f \approx 10 T$. 
In principle other ballistic channels with different
winding numbers might be present. Here they were too weak
to be detected.
%%%%%%%%%%%%%%%%%%%%%%%%%%%%%%%%%%%%%%%%%%
%%%%%%%%%%%% FIG 2 %%%%%%%%%%%%%%%%%%%%%%%
%%%%%%%%%%%%%%%%%%%%%%%%%%%%%%%%%%%%%%%%%%
\begin{figure}[htb]
\vspace{10pt}
\centerline{
\epsfig{file=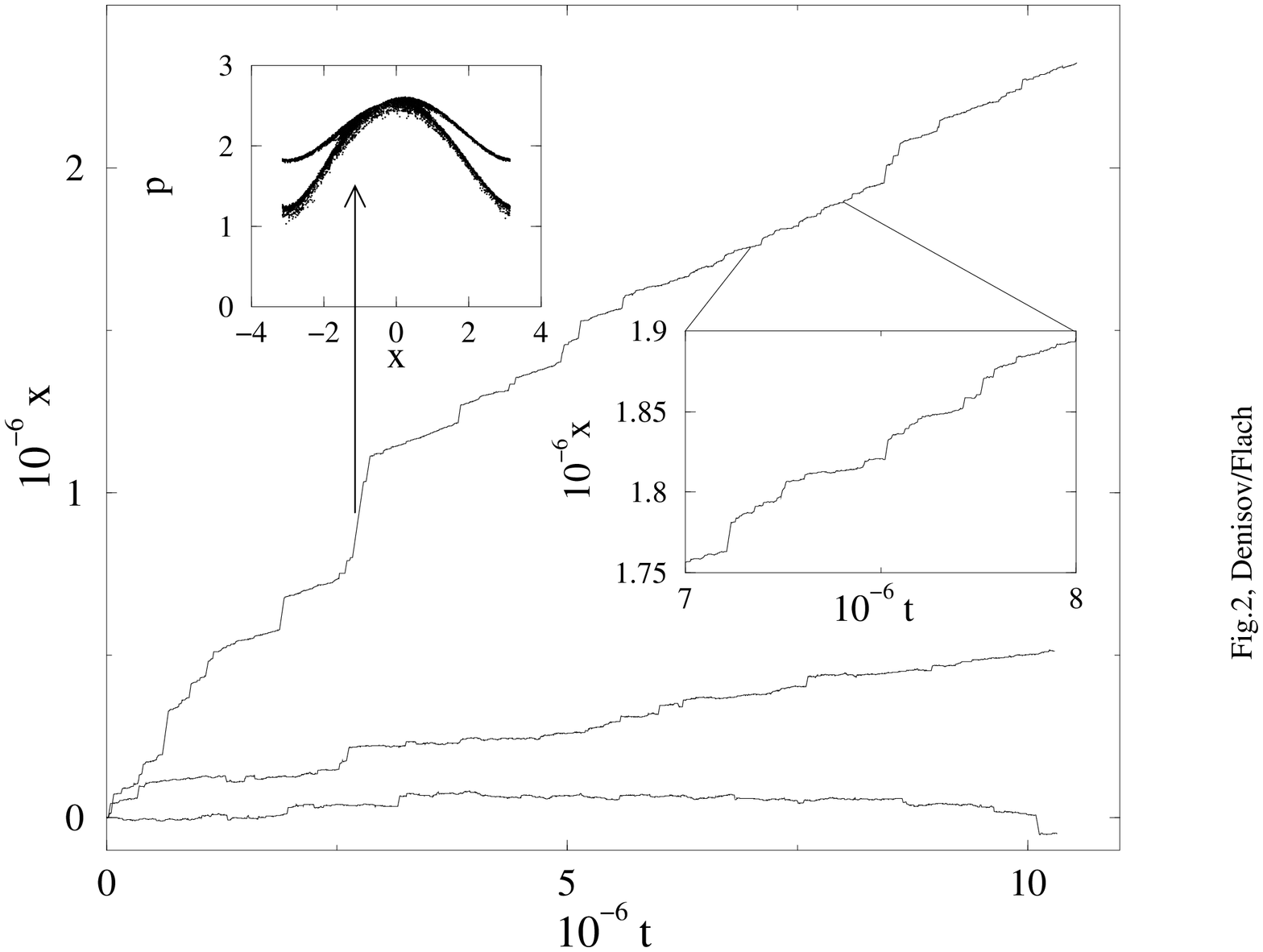,width=62mm,height=90mm,angle=270}
}
\vspace{2pt}
\caption{$x(t)$ for $\phi=0,\pi/5,\pi/2$ (lower, middle and upper
curves respectively). Left upper inset: Poincare map for ballistic
flight with $\phi=\pi/2$ as indicated by arrow. Right inset: zoom
of $x(t)$ for $\phi=\pi/2$.}
\label{fig2}
\end{figure}
%%%%%%%%%%%%%%%%%%%%%%%%%%%%%%%%%%%%%%%%%%%

In Fig.3(a) we show the corresponding distribution functions
$P_{\pm}(t_f)$ (again for $\phi=\pi/2$). They are obtained by counting 
the number of flights with $t_f$ falling into a time window of size $T$.
For $t_f < 10 T$ we observe exponential
dependence of $P_{\pm}$ on $t_f$. These short flight distributions
are in fact independent of the direction of flight.
Skipping all longer flights would lead to the prediction
of nearly zero average velocity (restricting to flights of length
$t_f < 10T$ yields about one percent of the numerically observed
current). 
Thus the desymmetrization will manifest itself for longer
flights.
For $t_f > 10 T$ a crossover to
a power law $P_{\pm} \sim t_f^{\alpha_{\pm}}$ takes place. Here we find
a significant desymmetrization for $\phi=\pi/5,\pi/2$. Estimating the
exponents \cite{num_errors}
we find for $\phi=\pi/5$: $\alpha_-\approx 2.5$,
$\alpha_+\approx 2.4$ and for $\phi=\pi/2$: $\alpha_-\approx 3.7$,
$\alpha_+\approx 2.3$. It is worthwhile noting that $\alpha < 3$
implies unidirectional anomalous diffusion with diverging second moments
of $P(t)$. The flights are coined {\it Levy flights} in such a case
\cite{Levy}.

Following the continuous-time random
walk (CTRW) formalism \cite{Klafter} we propose
a generalized asymmetrical flight model capable of reproducing
the above results.
The applicability of the CTRW model follows from the
assumption that the presence of a random phase 
with fast decaying correlations leads
to the absence of
correlations between consecutive flights, since they are 
almost always
separated by
dispersive chaotic motion.
%%%%%%%%%%%%%%%%%%%%%%%%%%%%%%%%%%%%%%%%%%
%%%%%%%%%%%% FIG 3 %%%%%%%%%%%%%%%%%%%%%%%
%%%%%%%%%%%%%%%%%%%%%%%%%%%%%%%%%%%%%%%%%%
\begin{figure}[htb]
\vspace{10pt}
\centerline{
\epsfig{file=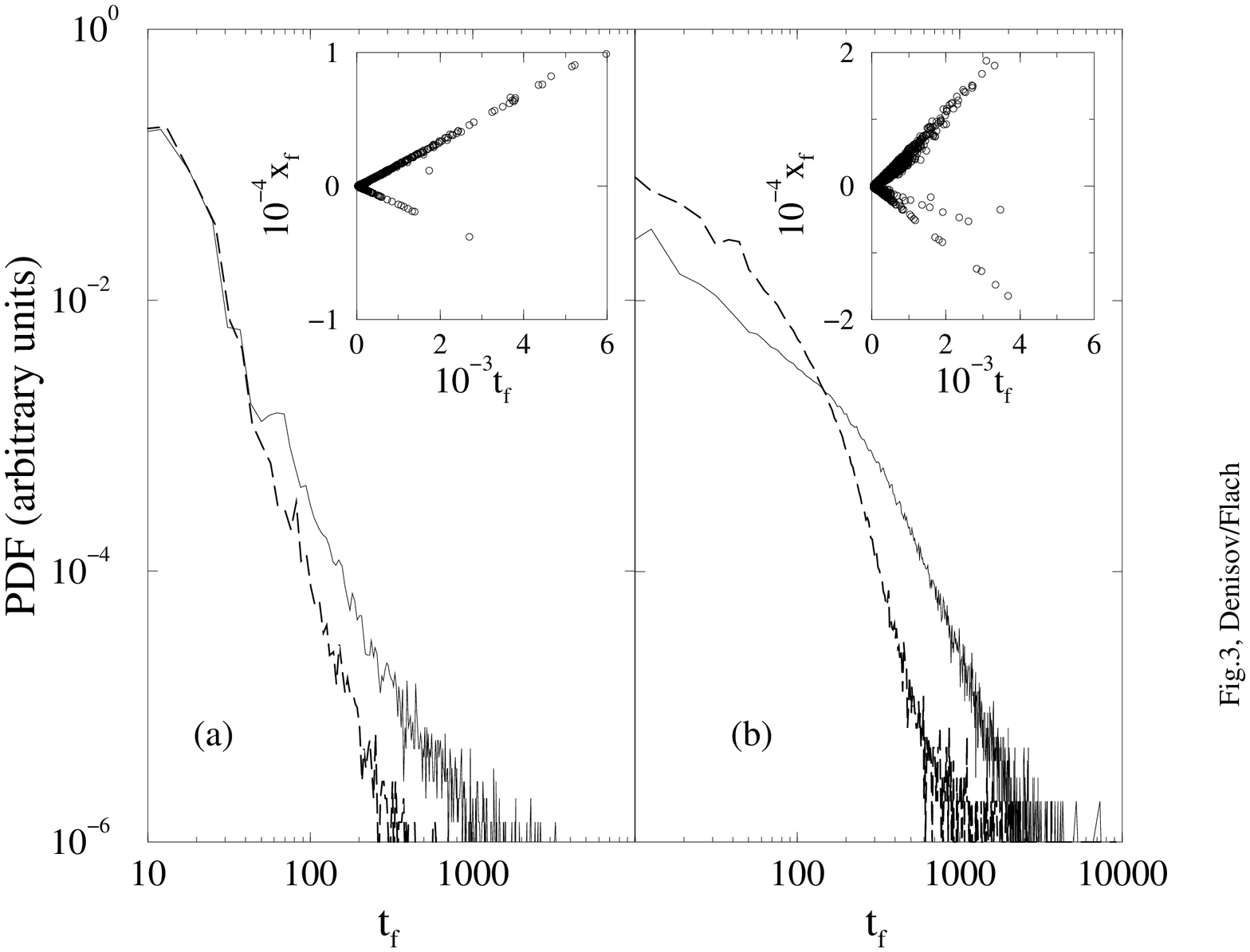,width=62mm,height=90mm,angle=270}
}
\vspace{2pt}
\caption{$P_+(t_f)$ (solid line) and $P_-(t_f)$ (thick dashed line).
Insets: $x_f$ versus $t_f$. For parameters see text.}
\label{fig3}
\end{figure}
%%%%%%%%%%%%%%%%%%%%%%%%%%%%%%%%%%%%%%%%%%%

Assume that there 
exist $N$ different  resonances with winding numbers
$w_i$, $i=1,...,N$. Every resonance is characterized by a probability
distribution function (PDF) of sticking time
$S_i(t)$.
After finishing a random phase event, the probability of sticking to
the $i$th resonance
is  $\rho_{i}$, $\sum_{i=1}^{N}\rho_{i}=1$.
The random phase residing time is characterized by a PDF 
$S_{r}(t)$.
All functions $S_{i}(t)$ and $S_{r}(t)$ must have finite first moments,
due to the Kac theorem about finiteness of recurrence times in Hamiltonian
systems \cite{Zas}. 
With these definitions we obtain 
the following expression for the current:
\begin{equation}
J=\frac{\sum_{i=1}^{N}\omega_{i}\rho_i\langle t_{i} \rangle}
{\sum_{i=1}^{N}\rho_i\langle t_{i} \rangle + \langle t_{r} \rangle}
\label{6}
\end{equation}
where $\langle t_i \rangle = \int t P_i(t) {\rm d}t$.

For the above discussed cases of $\phi=\pi/5,\pi/2$ we find
only two relevant ballistic channels - one with positive 
winding number and a second one with negative winding number.
In order to properly obtain $S(t)$, we note that our numerically
obtained function $P(t)$ consists of a lot of short 'flights' as
defined through the numerics \cite{com_num}. These may be either
stickings to islands with zero winding number or chaotic motion.
We observe that for flight times $t_f > 10 T$ only ballistic flights
with nonzero winding number are obtained. So the functions $S_{\pm}(t)$
may be easily obtained from $P_{\pm}(t_f)$ by cutting the central part
$t < 10 T$ out and properly normalizing. In this case the expression
for the average current simplifies to
\begin{equation}
J = \frac{1}{\kappa(1+f)}(\omega_+ \langle t_+ \rangle 
+ f \omega_-\langle t_- \rangle )
\label{jnew}
\end{equation}
where the two constants $\kappa$ and $f$ can be obtained from
the total time of a simulation $T_{tot}$ and the numbers $N_{\pm}$ of ballistic
flights, $\kappa = T_{tot} / (N_+ + N_-)$ and $f=N_-/N_+$.

For $\phi=\pi/5$ we obtain from the numerical runs
$f\approx 0.57$, $\kappa \approx 1900$, $\langle t_+\rangle \approx 
\langle t_- \rangle \approx 220$ and $\omega_+ =10/6 \approx 1.67$, 
$\omega_- = -1.5$. 
In this case of {\it weak desymmetrization} the main source of
a nonzero current is the different probability to enter a right or
left going flight because $f \neq 1$. At the same time the 
average flight times in both ballistic channels are nearly identical.
With the help of (\ref{jnew}) we find 
$J \approx 0.056$ which is close to the numerically observed value 0.05.

For the case $\phi=\pi/2$ we find
$f\approx 0.16$, $\kappa \approx 2600$, $\langle t_+\rangle \approx
400$, 
$\langle t_- \rangle \approx 150$ and $\omega_+ = 2$, 
$\omega_-\approx -1.4$. Note that the above discussed overlap with
the upper resonance yields a further {\it strong desymmetrization} in the
probability to realize a left or right going flight, and in addition
the average flight times in both channels significantly differ.
Expression (\ref{jnew}) yields $J\approx 0.22$ which is in good agreement
with the numerically observed value 0.2. 

For stronger driving amplitude $E_1=3.26$, $E_2=1.2$ and $\phi=\pi/2$
we obtain an average velocity $\langle \dot{x} \rangle \approx 0.85$.
The corresponding $x_f(t_f)$ dependence
and the PDFs $P_{\pm}(t_f)$ are shown in Fig. 3(b).
The $x_f(t_f)$ dependence shows that more than two ballistic channels
are involved.
The asymmetry of the PDFs at short flight times indicates
that a considerable number of left-going flights becomes
dominating at short times, 
in agreement with the tendency of the previous results.
This makes the application of the simplified sum rule (\ref{jnew}) 
impossible, instead the original definition (\ref{6}) should be
used. Careful analysis of the structure of the stochastic layer
shows that relevant resonances become embedded in the bulk of the
stochastic layer. While these structures are of rather small size,
they are frequently visited. A restriction to short flights $t_f <
10 T$ now yields a considerable nonzero current which is however
{\it negative}, i.e. opposite to the total current value. Again
the long ballistic flights are necessary in order to properly
obtain the observed current value.

In summary, we have explained the dynamical 
mechanisms of current appearance in
driven Hamiltonian systems inside the stochastic layer
with broken time-reversal symmetry. The
key
source of such a directed transport  is 
the desymmetrization of flight probabilities in ballistic channels
inside the layer. 
As it follows from our sum rule (\ref{6}) the resulting current
depends among other parameters on the average time spent in a 
ballistic channel. These times will sensitively depend on
control parameters of the system if the exponent $\alpha$ becomes
less than 3. In such a case small changes may significantly alter
the current value as shown above. 

A recently proposed geometric approach of counting areas
and winding numbers is in principle also capable of obtaining
the observed mean value for the current \cite{sumrule}.
This approach may also require sophisticated studies of the fractal
structure of the chaotic layer. It represents
a nontrivial complementary result, since it, although not
explaining the dynamical mechanisms of current generation,
is capable of obtaining the average current value (provided the
sums in Eq.(3) of \cite{sumrule} converge fast enough).

We thank M. Fistul and A. A. Ovchinnikov for useful discussions.


\begin{thebibliography}{99}
\bibitem{com_poinc}
The starting time of output was set to zero, subsequent points
are drawn after multiples of the drive period. The value of $x$
is mod $2\pi$, so the true trajectory may actually cover
large distances in $x$.
\bibitem{com_num}
Flights are monitored on a time grid with spacing $T$.
A flight end is defined by either a change of the velocity sign
or a shift in $x$ in the opposite direction.
\bibitem{num_errors}
Here and in the following our estimated numbers from fitting numerical
data are meant with a relative error of at least 10 percent.
\end{thebibliography}
\end{document}